\documentclass[conference, letterpaper, 10pt]{IEEEtran}
\IEEEoverridecommandlockouts
\usepackage{cite}
\usepackage{amsmath,amssymb,amsfonts}
\usepackage{algorithmic}
\usepackage{graphicx}
\usepackage{textcomp}
\usepackage{xcolor}
\usepackage{array}
\usepackage{amsmath}
\usepackage{url}
\usepackage{float}
\usepackage{subcaption}
\usepackage{booktabs}
\usepackage{color}
\usepackage{pifont}
\usepackage{booktabs}
\usepackage{enumitem}
\usepackage{multirow}
\usepackage{colortbl}
\usepackage[]{hyperref}
\usepackage{etoolbox}
\usepackage{tikz}
\usepackage[compact]{titlesec}

\usepackage{setspace}
\renewcommand{\thesubsubsection}{\Alph{subsection}.\arabic{subsubsection}}

\titleformat{\subsubsection}
  {\normalfont\itshape} 
  {\thesubsubsection} 
  {1em} 
  {} 

\titlespacing*{\subsubsection}
  {0pt} 
  {6pt} 
  {3pt} 

\usepackage[moderate, tracking=normal, title=normal, margins=tight, bibliography=tight, lists=normal, floats=tight, paragraphs=tight, mathspacing=normal, wordspacing=tight, leading=normal]{savetrees} 

\def\BibTeX{{\rm B\kern-.05em{\sc i\kern-.025em b}\kern-.08em
    T\kern-.1667em\lower.7ex\hbox{E}\kern-.125emX}}

\begin{document}

\title{\huge\textsc{Cryptonite}: Scalable Accelerator Design for Cryptographic Primitives and Algorithms
\vspace{-10pt}
}

\author{
\IEEEauthorblockN{Karthikeya Sharma Maheswaran\textsuperscript{\ddag}, 
                  Camille Bossut\textsuperscript{\textdagger}, 
                  Andy Wanna\textsuperscript{\ddag}, 
                  Qirun Zhang\textsuperscript{\textdagger}, 
                  Cong Hao\textsuperscript{\ddag}}
                  
\IEEEauthorblockA{\textsuperscript{\ddag}School of Electrical and Computer Engineering, Georgia Institute of Technology, USA \\ 
\textsuperscript{\textdagger}School of Computer Science, Georgia Institute of Technology, USA}
\{km304, cbossut21, awanna3, qrzhang, callie.hao\}@gatech.edu
}

\newcommand{\necessary}[1]{\textcolor{red}{#1}}
\newcommand{\camrevise}[1]{\textcolor{blue}{#1}}

\maketitle

\begin{abstract}

Cryptographic primitives, consisting of repetitive operations with different inputs, are typically implemented using straight-line C code due to traditional execution on CPUs. Computing these primitives is necessary for secure communication; thus, dedicated hardware accelerators are required in resource and latency-constrained environments. High-Level Synthesis (HLS) generates hardware from high-level implementations in languages like C, enabling the rapid prototyping and evaluation of designs, leading to its prominent use in developing dedicated hardware accelerators. However, directly synthesizing the straight-line C implementations of cryptographic primitives can lead to large hardware designs with excessive resource usage or suboptimal performance.

We introduce Cryptonite, a tool that automatically generates efficient, synthesizable, and correct-by-design hardware accelerators for cryptographic primitives directly from straight-line C code. Cryptonite first identifies high-level hardware constructs through verified rewriting, emphasizing resource reuse. The second stage automatically explores latency-oriented implementations of the compact design. This enables the flexible scaling of a particular accelerator to meet the hardware requirements. We demonstrate Cryptonite's effectiveness using implementations from the Fiat Cryptography project, a library of verified and auto-generated cryptographic primitives for elliptic-curve cryptography. Our results show that Cryptonite achieves scalable designs with up to 88.88\% reduced resource usage and a 54.31\% improvement in latency compared to naively synthesized designs.
\end{abstract}

\begin{IEEEkeywords}
High-level synthesis, Cryptography, Hardware acceleration, Cryptonite, \& Fiat Cryptography Project
\end{IEEEkeywords}

\section{Introduction}

Modern day computer security heavily relies on cryptography for authentication, encryption, and key exchange protocols. The implementation of cryptographic primitives, the foundational building blocks of cryptographic protocols, must be both correct and efficient to safeguard security and privacy. Improving the efficiency of computing such primitives is imperative, as it directly impacts the performance and scalability of security systems. Hardware acceleration has become a widely adopted methodology to meet this need, alleviating computational bottlenecks, enabling real-time processing, and reducing the risk of system vulnerabilities.

Ongoing efforts have prioritized manually accelerating cryptographic algorithms such as Number Theoretic Transforms (NTT)~\cite{10.1007/978-3-030-78713-4_6} and Polynomial Multiplications~\cite{9603378}, with an emphasis on creating efficient data-flow architectures \cite{chelton08}. Implementing such architectures is challenging, time-consuming, and requires expert knowledge. Therefore, there is a need to bridge the gap by automating the generation of such architectures.

Cryptographic primitives, usually written in C~\cite{10.1145/3611643.3613096}, contain repetitive computations to encrypt, decrypt, or sign using a particular scheme~\cite{eccapps11} with different inputs, making them ideal candidates for hardware acceleration. Being written in C with no high-level constructs, they can easily be synthesized using High-Level Synthesis (HLS). These straight-line codes lack branch statements, loops, and array constructs, leading to low-latency execution with excessive resource usage. While resource constraints can be applied using pragmas such as \texttt{\#pragma HLS allocation}, this usually results in complex control logic and bad timing~\cite{wu2021ironman,wu2022ironman}. Therefore, cryptographic primitives need to be rewritten for synthesis by HLS to generate high-performing, resource-efficient accelerators.

Cryptographic primitives are often deployed in resource-constrained environments~\cite{sandoval21, laranino20}, making it essential for synthesized designs to be \textbf{scalable}. This scalability enables the exploration of trade-offs between resource usage and latency. Real-world applications demand parallel execution of multiple primitives and co-optimization across them. Tuning the performance of each primitive is essential to achieve overall resource efficiency and performance optimization. 



For a given C program, the Vitis HLS compiler emits a single design. This design changes with different code structures, even if they are functionally equivalent. Exploring the resource-latency trade-off of cryptographic accelerators directly generated by HLS is challenging, as the straight-line C programs have no explicitly parameterizable constructs. Despite this, repeated patterns in straight-line C-code can be rolled into loop constructs, enabling design space exploration (DSE). This work tackles this challenge with automated, verified source-code rewriting, enabling architectural DSE. Users then choose from the Pareto-optimal designs to best meet hardware requirements. 

We target straight-line C programs from the Fiat Cryptography project \cite{fiat19}, a library of verified and auto-generated cryptographic primitives for elliptic-curve cryptography. Using two implementation baselines, one with the original straight-line code from the Fiat Project and the other with loop constructs, we reduced resource utilization and comparable latency to the baselines. 

This work, Cryptonite, has three main contributions:
\begin{itemize}
    \item Verified source-to-source HLS transformations, generating array and loop constructs from straight-line code using equivalence graphs (e-graphs).
    \item Automatic Design Space Exploration over looped code, resulting in a range of previously unattainable points with varying resource-latency profiles.
    \item Exploring the co-optimization of multiple primitives with parallel execution for scalable accelerator design.
\end{itemize}


The rest of the paper is organized as follows. \hyperref[sec:motivation]{Section~\ref*{sec:motivation}} frames and motivates this work. \hyperref[sec:Approach]{Section~\ref*{sec:Approach}} provides an overview of the \textbf{Cryptonite}\footnote{This tool has been open-sourced at \url{https://github.com/KarthikeyaSharma16/Cryptonite}} tool. \hyperref[sec:Results]{Section~\ref*{sec:Results}} presents our experimental results, and \hyperref[sec:RelatedWork]{Section~\ref*{sec:RelatedWork}} summarizes and addresses related work. Finally, \hyperref[sec:Conclusion]{Section~\ref*{sec:Conclusion}} concludes the paper.

\section{Motivation} \label{sec:motivation}

\begin{figure*}[h!]
    \centering
    \includegraphics[width=1.05\linewidth]{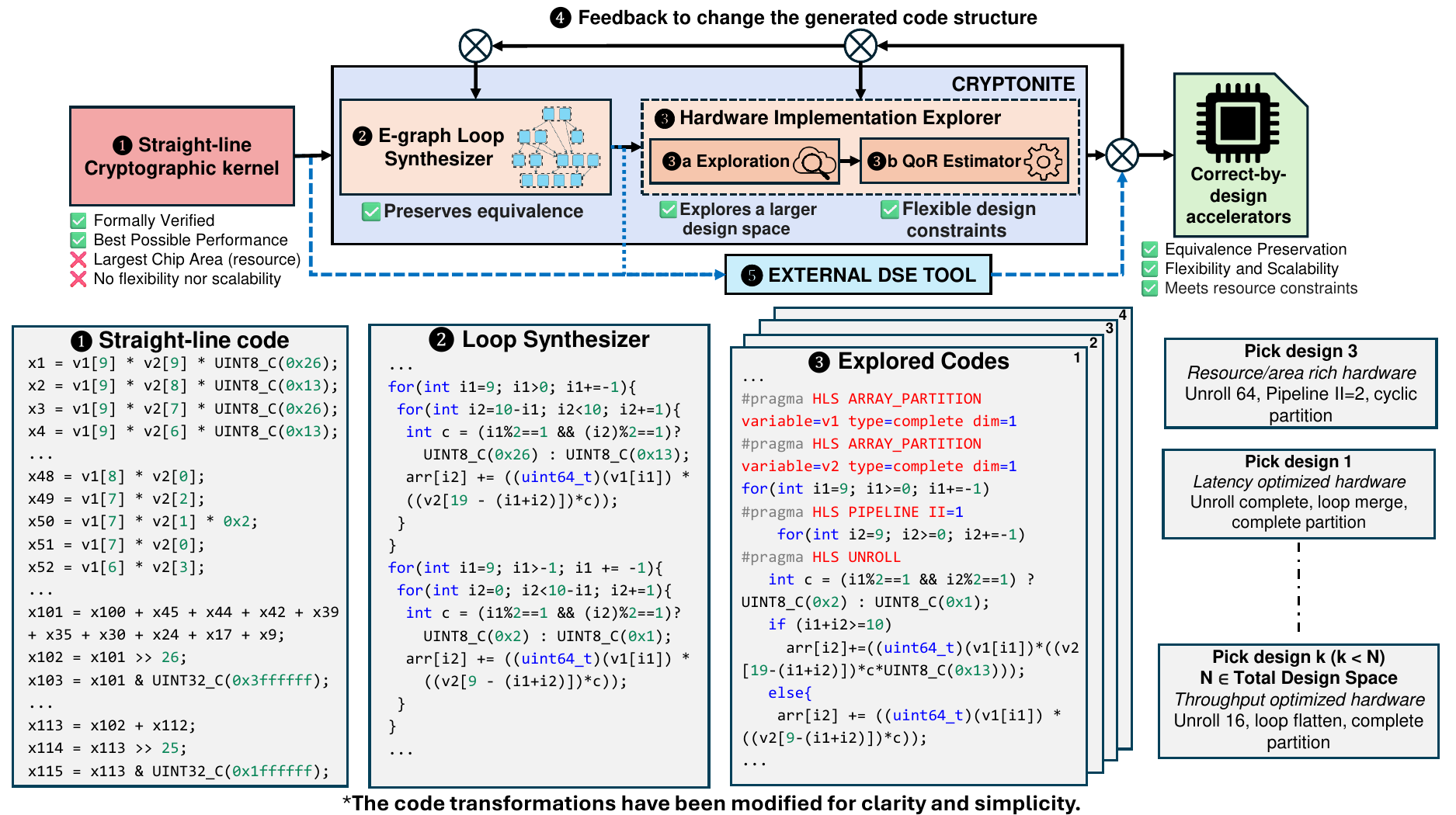}
    \caption{Cryptonite workflow.}
    \label{fig:blockdiagram}
    \vspace{-15pt}
\end{figure*}

The advent of hardware accelerators has optimized compute-intensive tasks such as Machine Learning (ML) algorithms by leveraging structured code with high data reuse, minimal dependencies, and predictable memory access patterns, enabling efficient parallelization based on workload demands. Cryptographic primitives, on the other hand, have different memory access patterns from ML code that are also more irregular. These are often straight-line programs that prioritize runtime efficiency rather than decipherable code structure. \hyperref[fig:blockdiagram]{Fig.~\ref*{fig:blockdiagram}} illustrates an example of a straight-line program (\ding{182}) with a dependency structure. Note that the displayed code is a simplified version of the original code. In this code, no clear patterns emerge, and redundant computations are repeated across different code segments. The non-trivial data dependencies due to complex arithmetic, irregular array access patterns, and the absence of loops result in an inefficient algorithm-to-hardware mapping. For instance, if a straight-line code contains 100 independent assignment operations involving at least one 128-bit multiplication, DSP resources are statically allocated for every multiplication. This results in poor resource reuse and limits scalability for larger cryptographic workloads. To quantify the inefficiencies on hardware utilization and scalability, we evaluate the performance of this workload using a metric called \textbf{\textit{Normalized Performance Index (NPI)}}, which is defined in \hyperref[eq:npi]{Eq.~\ref*{eq:npi}}: 
\vspace{0.005\linewidth}

\begin{equation} \label{eq:npi} 
\mathit{NPI} = w_1 \frac{\mathit{l} - \mathit{l}_{\min}}{\mathit{l}_{\max} - \mathit{l}_{\min}} + w_2 \frac{\mathit{r} - \mathit{r}_{\min}}{\mathit{r}_{\max} - \mathit{r}_{\min}}
\vspace{0.03\linewidth}
\end{equation}

\noindent Where $l$ is the latency of a design and $r$ is the percentage of resources used. In this section, the NPI is computed with $w_1 = w_2 = 0.5.$ A lower NPI metric indicates a more optimal balance between resource utilization and latency. The percentage of resources used $r$ is determined with \hyperref[eq:r_used]{Eq.~\ref*{eq:r_used}}: 

\begin{equation} \label{eq:r_used}
\mathit{r} = \frac{1}{4} \left( 
\frac{\mathit{DSP}_{\mathit{u}}}{\mathit{DSP}_{\mathit{t}}} + 
\frac{\mathit{LUT}_{\mathit{u}}}{\mathit{LUT}_{\mathit{t}}} + 
\frac{\mathit{FF}_{\mathit{u}}}{\mathit{FF}_{\mathit{t}}} + 
\frac{\mathit{BRAM}_{\mathit{u}}}{\mathit{BRAM}_{\mathit{t}}} 
\right) \times 100
\end{equation}
\vspace{0.01\linewidth}

\hyperref[tab:motivation_benchmarks]{Table~\ref*{tab:motivation_benchmarks}} summarizes the results obtained after testing two cryptographic kernels across multiple configurations, starting with the baseline code. Amongst all the tested cases, the baseline kernels exhibit the highest NPI values, indicating suboptimal performance. A naive approach to maximizing resource usage through built-in pragma (\texttt{\#pragma HLS allocation}) led to poor hardware mapping and increased design latency, underscoring the need for a more sophisticated solution. Testing the baseline code with state-of-the-art (SOTA) DSE tools for HLS, such as ScaleHLS, showed negligible improvement over the baseline, highlighting the challenges of optimizing straight-line cryptographic workloads. Introducing loop structures into the baseline code improved hardware mapping, but at the cost of higher latency. However, when this code was optimized using ScaleHLS, the NPI decreased further, yielding better performance. Our proposed DSE framework yielded the most significant improvements, achieving the lowest NPI values -- kernel \#1 demonstrated a 3.5$\times$ performance gain, while kernel \#2 achieved 4.5$\times$ improvement.

\begin{table}[h!]
\centering
\small
\setlength{\tabcolsep}{3pt}
\begin{tabular}{
    >{\centering\arraybackslash}p{17em} | 
    >{\centering\arraybackslash}p{4.5em} | 
    >{\centering\arraybackslash}p{4.5em}
}
    \toprule
    \textbf{Cases $\downarrow$ / Benchmarks $\rightarrow$} & \textbf{NPI for Kernel \#1} & \textbf{NPI for Kernel \#2} \\
    \hline
    Baseline \ding{182} & 0.5 & 0.499 \\
    \hline
    Baseline + Allocation Pragma & 0.5 & 0.5 \\
    \hline
    Baseline \ding{182} + ScaleHLS \ding{182} + \ding{186} & 0.5 & 0.4903  \\
    \hline
    Looped code \ding{182} + \ding{183} & 0.4705 & 0.35 \\
    \hline
    Looped code + ScaleHLS \ding{182} + \ding{183} + \ding{186} & 0.3496 & 0.179 \\
    \hline
    Baseline \ding{182} + Cryptonite \ding{183} + \ding{184} + \ding{185} & \textbf{0.1398} & \textbf{0.111} \\
    \bottomrule
\end{tabular}
\caption{\centering Comparison of NPI for two benchmarks.}
\label{tab:motivation_benchmarks}
\vspace{-5pt}
\end{table}

Thus, straight-line code is a poor target for high-level DSE as it lacks \textbf{\textit{parametrizable constructs}}. SOTA DSE tools~\cite{ye2021scalehls, 10.1145/3572959} primarily rely on source-code loop transformations, such as tiling, padding, and perfectization, along with HLS pragmas that influence loop behavior to explore various resource-latency trade-offs. However, in the absence of loop structures, these techniques become ineffective. This makes scalable designs harder to synthesize. Resource limitations are typical in the domain of cryptography, as encryption is still necessary in constrained environments. Thus, re-discovering parametrizable constructs in straight-line code is needed to explore a broader resource/latency tradeoff, imperative to the practical deployment of cryptographic kernels. 

Looping straight-line code generally enables improved hardware mapping through regular memory access patterns and resource scheduling. Eliminating or minimizing data dependencies and identifying repeated computation patterns enables time-multiplexed DSP usage instead of static allocation. By systematically restructuring the looped code, one can explore various trade-offs and benefits, ultimately designing a highly efficient architecture. \hyperref[fig:blockdiagram]{Fig.~\ref*{fig:blockdiagram}} presents an example of a structured code (\ding{183}) that also has straight-line code lacking loop structures. Such straight-line code structures make it challenging to incorporate loops and require extensive exploration of design choices to understand their impact on performance and guide the next phase of design iteration. Exploring such design choices for the cryptographic kernels has two key challenges: 

\ding{172} \textbf{Formal guarantee of transformations}. Correctness is paramount to cryptographic applications -- any unintended transformation could introduce vulnerabilities, breaking security guarantees. Conventional DSE for ML workloads allows heuristic-driven optimizations since floating-point approximations (quantization and pruning) are often acceptable due to the robustness of the model, whereas cryptographic workloads require provably correct transformations that are bit-accurate to maintain data integrity.

\ding{173} \textbf{Looping the auto-generated straight-line code}. Cryptographic primitives use a combination of modular arithmetic, bitwise shifts, and rotations with operands that are not regularly placed in memory. This makes it difficult to automatically structure the source code to enable resource reuse for logic synthesis. To explore the design space for hardware implementation of these primitives, we must first identify patterns in the code and reconstruct computationally efficient loop structures while preserving formal correctness.


These challenges call for a specialized tool that expands DSE's reach in cryptography by reintroducing parametrizable code structures into straight-line code.



\section{Approach}
\label{sec:Approach}


\hyperref[fig:blockdiagram]{Fig.~\ref*{fig:blockdiagram}} presents the workflow of our DSE tool, Cryptonite. Cryptonite begins with straight-line primitives as inputs and performs DSE in two stages -- \ding{182} \textbf{E-graph Loop Synthesizer (\hyperref[subsec:E-graph_Loop_Synth]{Sec.~\ref*{subsec:E-graph_Loop_Synth}})} and \ding{183} \textbf{Hardware Implementation Explorer (\hyperref[subsec:Hardware_Impl_Explorer]{Sec.~\ref*{subsec:Hardware_Impl_Explorer}})}. In the first stage, the E-graph loop synthesizer transforms local variables into arrays to perform Loop Synthesis. The resultant structured code is then passed to the Hardware Explorer which evaluates optimization strategies based on QoR estimator to enhance performance. \hyperref[fig:blockdiagram]{Fig.~\ref*{fig:blockdiagram}\ding{186}} demonstrates the potential for integration of any external DSE frameworks that can make use of the output of the E-graph Loop Synthesizer (or) the straight-line cryptographic kernel.

\subsection{E-graph Loop Synthesizer}
\label{subsec:E-graph_Loop_Synth}
\begin{figure}[!t]
    \centering
    \includegraphics[height=1.2\linewidth]{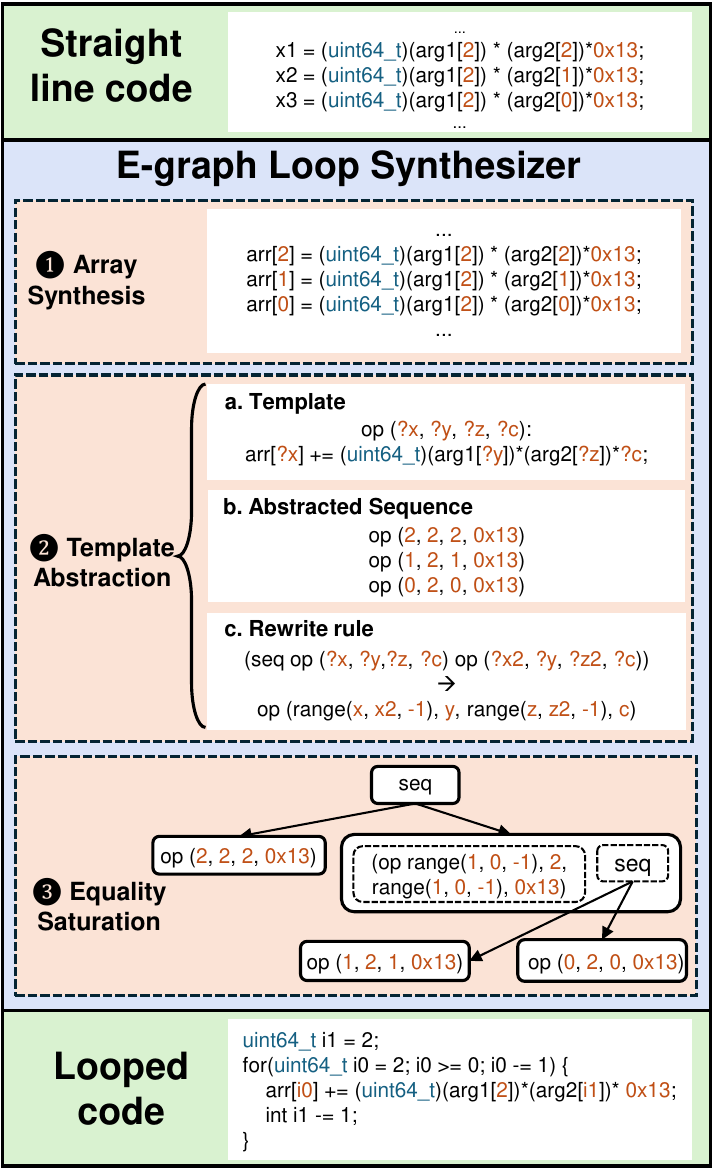}
    \caption{E-graph Loop Synthesizer.}
    \label{fig:egg_looping}
\end{figure}

Local variables in straight-line primitives are replaced with array elements so that they can be indexed with constant values. This change helps to eliminate excess registers that HLS could not optimize. It also opens up opportunities to inject loops, as we can now reference the variables with constant indices expressed in terms of loop variables. However, the challenge is that replacing multiple local variables with one array element can introduce data dependencies. To address this, we construct a data-dependence graph (DDG) from the input code using the SVF framework\cite{SuiX16svf}. We generate a value-flow graph for our input program and perform an analysis of this graph, which identifies local variables that are used once only after assignment. Such variables consumed by the same expression can safely be collapsed into a single array index without affecting code correctness. We also track the variable types and only group those with the same type into the same array and/or element. Once variables are assigned to an array element, we choose concrete array indices for each group and replace them accordingly in the source code. While assigning indices arbitrarily would already reduce the number of local variables, our approach goes further: we base the assignment on access patterns of other input arrays in the targeted sections of code to enhance future loop injection opportunities. \hyperref[fig:egg_looping]{Fig.~\ref*{fig:egg_looping}} gives an example of straight-line C-code and the corresponding code with the synthesized arrays. We identify two essential code properties for synthesizing loops from straight-line code: 

\ding{182} \textbf{\textit{Code patterns.}} To inject loops, we need repetitive code. However, syntactic differences between lines of code can obscure semantic patterns. To address this, we abstractly interpret sequential variable assignments in the clang AST as mathematical formulas using a C++ library called GiNaC~\cite{ginac}. Array elements used in the lines of code are abstracted as mathematical variables, where the same array corresponds to the same variable. If the mathematical structures of two sequential lines match, we take note of the formula as a frequently used pattern. 

\ding{183} \textbf{\textit{Index dependence.}} Each variable in the target code section must be expressible in terms of a constant value. The purpose of this is to ensure that all lines of code can be expressed in terms of a C-code template and a loop variable (which will replace the constant values in the synthesized loop). To achieve this, the variables present must be array elements with constant indices, or the same variable used repeatedly.

We sweep the target function for lines of code with the properties above. For these lines, we construct a C-code \emph{template} for the structure and a symbolic equation for the mathematical semantics of the line. Due to property \ding{183}, the holes in the C-code template will always be filled with constant values in the target line. Thus, we can represent the line as a pair formed by its C-code template and the set of constants to complete it. Using this templating method, we group lines of code for which the mathematical semantics match, meaning the same C-code template can represent them. The abstracted version of the straight-line C-code in \hyperref[fig:egg_looping]{Fig.~\ref*{fig:egg_looping}} is under abstracted Sequence with the corresponding template above. Once we have considered all lines of the function, we output all sets of constants that correspond to the same template along with their line numbers. This forms sequences of abstract operations with constant arguments.

The purpose of this template abstraction is to more easily construct an e-graph representation of the code. E-graphs are data structures that can compactly represent equivalent versions of an expression or program~\cite{egraphs09}. An e-graph is expanded to represent more equivalent versions by applying rewrite rules. Applying a set of rewrite rules exhaustively to an e-graph is called equality saturation. We discover loops using equality saturation over our e-graph representation of the code. This is the heart of \emph{loop saturation}. Our approach uses \textit{egg} \cite{WillseyNWFTP21egg}, a fast and extensible e-graph framework for equality saturation. 

To obtain looped programs, we need rewrite rules that introduce loops in our e-graph. We construct these rules using the corresponding C-code template and the program's data dependence graph. An intuitive way of introducing loops is to create a C-code-like for-loop operator containing a loop variable with a start, stop, and step. The issue with this approach is that it requires all constants in the abstracted operations to be expressed in terms of this variable. In cases where there are multiple constants that must be expressed in terms of a loop variable, like for the Abstracted Operations in \hyperref[fig:egg_looping]{Fig.~\ref*{fig:egg_looping}}, this means we must synthesize a linear equation that expresses the replaced constants in terms of the loop variable. This kind of equation synthesis is unnecessary for the scope of this project. Instead of expressing all variables in terms of one loop variable, which is time-consuming and restrictive, we insert loop-like structures \emph{within} an abstracted operation, which summarize the start, stop, and step conditions for that particular abstracted argument. That way, each argument can have its own potential loop variable. We call this operator a \emph{range} operator, as it describes a variable range. The rewrite rules that insert loops into the e-graph actually merge operations into a loop by replacing their operands with the range operator. The rewrite rule used for this transformation is shown in \hyperref[fig:egg_looping]{Fig.~\ref*{fig:egg_looping}} under part 2c (Rewrite Rule). An intermediate e-graph phase with one application of this rule is shown under Equality Saturation in the same diagram. The output looped code of the e-graph loop synthesizer is shown at the bottom of \hyperref[fig:egg_looping]{Fig.~\ref*{fig:egg_looping}}.

The e-graph loop synthesizer always produces the most looped version of the provided code. Therefore, we can only incorporate feedback by changing the code we input. We incorporate feedback from the final synthesis step by including and excluding target code sequences in the e-graph re-rolling step. In particular, we can exclude targeted code sequences that contain the fewest number of operations; even for the most compact loop representation, the overhead of representing the loop structure may not outweigh the DSE benefits it enables. This produces differently looped codes to input for the DSE step that follows.


The approach for equality saturation consists of the following steps: \ding{172} Inject loops with possible variable steps for pairs of instructions. \ding{173} Apply rewrite rules that absorb surrounding instructions into these proposed loops. \ding{174} Extract the best-looped version according to e-graph AST size. \ding{175} Use the C-code template for the e-graph operation to output C-code loops from the extracted e-graph.




\subsection{Hardware Implementation Explorer} 
\label{subsec:Hardware_Impl_Explorer}

\begin{table*}[!t]
\centering
\small

{
\resizebox{\textwidth}{!}{%
\begin{tabular}{
|
>{\centering\arraybackslash}m{1.6cm}|
>{\centering\arraybackslash}m{2.9cm}|
>{\centering\arraybackslash}m{1.0cm}
>{\centering\arraybackslash}m{1.8cm}
>{\centering\arraybackslash}m{1.8cm}
>{\centering\arraybackslash}m{1.0cm}
>{\centering\arraybackslash}m{1.0cm}|
>{\centering\arraybackslash\columncolor[gray]{0.9}}m{1.0cm}
>{\centering\arraybackslash\columncolor[gray]{0.9}}m{1.4cm}
>{\centering\arraybackslash\columncolor[gray]{0.9}}m{1.8cm}
>{\centering\arraybackslash\columncolor[gray]{0.9}}m{1.0cm}
>{\centering\arraybackslash\columncolor[gray]{0.9}}m{1.0cm}|
}

\hline
\textbf{Primitive} & \textbf{Design Approach} & \textbf{Latency (cycles)} & \textbf{DSP (\%)} & \textbf{LUT (\%)} & \textbf{NPI} & \textbf{Speedup} & \textbf{Latency (cycles)} & \textbf{DSP (\%)} & \textbf{LUT (\%)} & \textbf{NPI} & \textbf{Speedup}\\
\hline
\hline

\multirow{8}{1.6cm}{\centering \texttt{Curve 25519}} & & & & &  &  &  & & & &\\
& Original Code  & 95 & 400 (15.87) & 12414 (4.529) & 0.5776 & 1$\times$ & 116 & 220 (8.73) & 7165 (2.614) & 0.84 & 1$\times$\\
 & Rolled Code  &  270 & 10 (0.39) & 3065 (1.11) & 0.5 & 0.35$\times$& 148 & 102 (4.04)& 6051 (2.207) & 0.5 & 0.78$\times$ \\
& Latency Optimized $\ast$  & 67 & 428 (16.98)& 11921 (4.34) & 0.536 & 1.42$\times$  & 52 & 174 (6.9)& 6776 (1.472) & 0.3407 & 2.23$\times$ \\
  & Resource Optimized $\ast$  & 95 & 40 (1.58)& 4827 (1.76) & 0.149 & 1$\times$  & 88 & 104 (4.12)& 8537 (3.114) & 0.336 & 1.17$\times$ \\
 & Latency Optimized $\dagger$& 51 & 428 (16.98) & 11716 (4.274) & 0.498 & 1.86$\times$ &  45 & 152 (6.03)& 7331 (2.674) & 0.242 & 2.57$\times$ \\
 & Resource Optimized $\dagger$& 231 & 14 (0.55)& 7023 (2.56) & 0.495 &  0.41$\times$ & 57 & 100 (3.96) & 7442 (2.715) & 0.109 & 2.03$\times$  \\
  & Scale-HLS & 127 & 50 (1.98)& 14832 (5.411)& 0.367 & 0.74$\times$ & 91 & 168 (6.67) & 10353 (3.77) & 0.69 & $1.27\times$\\
\hline
\hline

\multirow{8}{1.6cm}{\centering \texttt{Curve P521}} & & & & &   &   &  & &  & &\\
& Original Code  & 78 & 1296 (51.42) & 23227 (8.474) & 0.581 & 1$\times$ & 84 & 760 (30.15) & 20798 (7.588) & 0.51 & 1$\times$\\
 & Rolled Code & 155 & 288 (11.4)  & 7186 (2.621) & 0.554 & 0.5$\times$ & 245 & 416 (16.5) & 16973 (6.192) & 0.618 & 0.34$\times$\\
& Latency Optimized $\ast$ & 70 & 752 (29.8)  & 10312 (3.76) & 0.271 & 1.11$\times$ & 79 & 424 (16.82)  & 24039 (8.77)& 0.209 &1.06$\times$\\
  & Resource Optimized $\ast$ & 105 & 144 (5.71) & 10159 (3.7) & 0.232 & 0.74$\times$ & 179 & 272 (10.79) & 18328 (6.687) & 0.307 & 0.46$\times$ \\
 & Latency Optimized $\dagger$ & 63 & 1296 (51)  & 18653 (6.80) & 0.485 & 1.23$\times$  & 103 & 308 (12.22)  & 19544 (7.13) &  0.117 & 0.81$\times$ \\
 & Resource Optimized $\dagger$ & 89 & 144 (5.71)  & 9246 (3.37) & 0.141 & 0.87$\times$  & 110 & 280 (11.11) & 17846 (6.511) & 0.093 &  0.76$\times$   \\
  & Scale-HLS & 104 & 128 (5.079) & 16043 (5.85) & 0.262 & 0.75$\times$ &  - & - & - & - & - \\
  
\hline
\hline



\multirow{8}{1.6cm}{\centering \texttt{Curve Secp256k1}} & & & & &   &  &   &  & & &\\
& Original Code  & 66 & 81 (3.21) & 12105 (4.41) & 0.5 & 1$\times$ & 64 & 44 (1.74) & 9292 (3.39) & 0.5 & 1$\times$\\
 & Rolled Code & 211 & 63 (2.5) & 6954 (2.53) & 0.67 & 0.31$\times$ & 306 & 7 (0.27) & 6954 (2.53) & 0.5 & 0.23$\times$\\
& Latency Optimized $\ast$ & 169 & 43 (1.7) & 7173 (2.61) & 0.468 & 0.39$\times$ & 240 & 19 (0.75)  & 6523 (2.37) & 0.47 & 0.18$\times$ \\
  & Resource Optimized $\ast$ & 182 & 39 (1.54) & 9067 (3.308) & 0.721 & 0.36$\times$ & 246 & 11 (0.43) & 8114 (2.96) & 0.78 & 0.26$\times$  \\
 & Latency Optimized $\dagger$ & 143 & 48 (1.9) & 9227 (3.366) & 0.54 & 0.48$\times$ & 118 & 32 (1.26) & 7441 (2.71) & 0.386 & 0.54$\times$ \\
 & Resource Optimized $\dagger$ & 190 & 32 (1.26)  & 5118 (1.86) & 0.427 & 0.34$\times$ & 275 & 7 (0.27)  & 7684 (2.8) & 0.504 &  0.26$\times$ \\
  & Scale-HLS & - & - & - & - & - & - & - & - & - & -  \\
  
\hline
\hline

\multirow{8}{1.6cm}{\centering \texttt{Curve P384}} & & & & &  &  &    &  & & &\\
& Original Code  & 80 & 112 (4.44) & 10182 (3.71) & 0.5 & 1$\times$ & 61 & 144 (5.71) & 10943 (3.99) & 0.5 &  1$\times$\\
 & Rolled Code & 598 & 32 (1.26)  & 11197 (4.085) & 0.717 & 0.13$\times$ & 425 & 32 (1.26) & 12113 (4.41) & 0.65 & 0.14$\times$ \\
& Latency Optimized $\ast$ & 107 & 96 (3.8) & 8948 (3.264) & 0.379 & 0.74$\times$ & 110 & 120 (4.76)  & 9651 (3.52) & 0.409  & 0.55$\times$ \\
  & Resource Optimized $\ast$ & 187 & 56 (2.22) &  13012 (4.747) & 0.576 & 0.42$\times$ & 106 & 56 (2.22) & 9488 (3.46) & 0.17 & 0.57$\times$  \\
 & Latency Optimized $\dagger$ & 133 & 75 (2.85) & 7284 (2.65) & 0.051 & 0.6$\times$ & 89 & 96 (3.8)  & 9689 (3.53) & 0.33 & 0.68$\times$  \\
 & Resource Optimized $\dagger$ & 145 & 64 (2.53)  & 9957 (3.63) & 0.37 & 0.55$\times$ & 124 & 40 (1.58)  & 8499 (3.10) & 0.086 & 0.49$\times$  \\
  & Scale-HLS & - & - & - & - &- &- & - & - &- &-  \\
  
\hline
\end{tabular}}}

\caption{\centering Comparison of resource-latency trade-offs for Manual DSE ($\ast$), Cryptonite ($\dagger$) and Scale-HLS. Resource-latency trade-offs are evaluated using either a Multiplication Kernel (white) or Square Kernel (grey).}
\label{tab:table}
\vspace{-15pt}
\end{table*}

Recall that straight-line code is a poor target for high-level DSE as it lacks parametrizable constructs. Synthesizing loop structures enables source-code transformations to explore designs with different performance characteristics. However, existing SOTA DSE tools fall short in our domain because they rely on MLIR, specifically the Affine and Structured Control Flow (SCF) Dialects~\cite{ye2021scalehls, seer23, 10.1145/3572959}. These can support loop transformations but restrict expressions to additions, subtractions, and multiplications. They further expect highly regular loop structures. Cryptographic algorithms, however, depend on non-affine operations like modulo, division, and bit-wise logic, which disrupt these transformations and introduce irregularities. While MLIR’s optimization infrastructure~\cite{lattner2020mlircompilerinfrastructureend} excels in ML and numerical workloads with predictable data access patterns, it is less effective for our needs. Without specialized dialects or extensions, these tools cannot fully leverage the parallelization opportunities in cryptographic workloads. Furthermore, the lack of open-source availability of DSE frameworks prevents us from evaluating their effectiveness on cryptographic workloads. To address this, we developed Cryptonite’s hardware implementation explorer, which incorporates a QoR estimator to rapidly synthesize scalable and efficient designs under varying constraint requirements. The explorer operates in two steps: Exploration, where design variants are generated, and QoR Estimation, where each variant is evaluated against performance and resource constraints.

\subsubsection{Exploration} The code transformations described below are applied directly to the e-graph synthesized code structure (\hyperref[fig:blockdiagram]{Fig.~\ref*{fig:blockdiagram}}\ding{183}), which is well-suited for such optimizations. The main stages of the explorer are as follows:

\ding{182} \textbf{Loop Pattern Analyzer.} This stage analyzes the structured output of the e-graph loop synthesizer to extract key features such as loop boundaries, loop-carried dependencies, array access patterns, loop nesting levels, and function calls. It also identifies and lifts redundant straight-line array accesses into loop structures based on detected index patterns. Sometimes, the overall code structure is also changed by creating \textit{reusable} functions for repeated code patterns with flow dependencies. Instead of duplicating the hardware for each recurring code pattern, the defined function is reused, thus reducing resource utilization.
    
\ding{183} \textbf{Variable bound removal.} This stage resolves variable loop bounds into static constants, as most loop transformations and HLS optimizations require statically bounded loops. Normalizing bounds also mitigates variability that can lead to side-channel vulnerabilities.

\ding{184} \textbf{Loop Exploration.} This stage generates diverse design variants that capture a wide range of performance trade-offs by systematically applying a combination of loop transformations. Rather than following a fixed transformation pipeline, the explorer enumerates and applies multiple combinations of optimizations to construct a rich design space. Each input loop nest is analyzed to determine valid transformation opportunities, and all permissible variants are explored. While the loop transformation flow follows a logical order starting with loop interchange and padding, followed by fusion, loop perfectization, branch elimination, strength reduction, and tiling, it is not enforced as a fixed pipeline. Instead, this order helps determine legal and beneficial transformation combinations. The explorer selectively applies subsets of these transformations to generate a diverse set of loop variants. \hyperref[fig:blockdiagram]{Fig.~\ref*{fig:blockdiagram}}\ding{184} shows an example of an explored code variant, where two nested loops are combined into a single loop through loop fusion. Loop padding is applied to align the iteration spaces, enabling a legal loop fusion output. The resulting structure allows for controlled parallelism supported by pipelining and loop unrolling directives.

    
\ding{185} \textbf{Pragma Insertions.} This stage analyzes the transformed loop structures to determine where and how pragma directives should be applied. These directives include, but are not limited to, array partitioning, loop unrolling, loop pipelining, loop flattening, and loop merging to improve parallelism, along with customizable parameters per directive applied. Additionally, dependence pragmas, leveraging dependency information extracted from code analysis, help resolve false hazards. Since each pragma directive includes configurable parameters, the explorer systematically generates all valid permutations of pragma configurations for each loop nest. It explores combinations of synthesis directives—such as array partitioning granularities, pipeline initiation intervals, and unroll factors—to produce a diverse set of implementations that span a wide range of performance and resource trade-offs.


\ding{186} \textbf{Design Space Synthesis.}  In this stage, the design variants generated by stage \ding{184} are combined with the pragma configurations identified during stage \ding{185}. While the Loop DSE stage produces a diverse set of loop transformation variants, the Pragma Insertion stage determines applicable directives and their parameters for each loop structure. These pragma annotations are stored separately as metadata and are merged with their corresponding loop variants to construct the final design space. The synthesized design space is the input for downstream QoR estimation, enabling guided selection of the most optimal implementation under given constraints.



\begin{figure*}[!htb]
    \centering
    \begin{subfigure}[t]{0.41\linewidth}
        \centering
        \includegraphics[width=\linewidth]{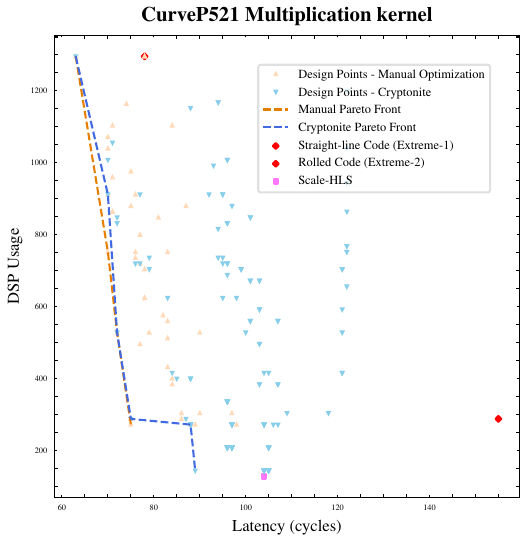}
        \caption{Design Points for CurveP521 Multiplication Kernel}
        \label{fig:paretoA}
    \end{subfigure}
    \hspace{0.05\linewidth} 
    \begin{subfigure}[t]{0.41\linewidth}
        \centering
        \includegraphics[width=\linewidth]{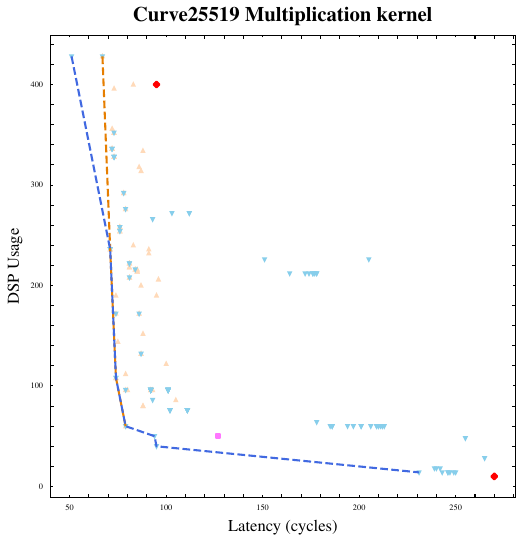}
        \caption{Design Points for Curve25519 Multiplication Kernel}
        \label{fig:paretoB}
    \end{subfigure}
    
    \vspace{0.5em} 
    
    \begin{subfigure}[t]{0.41\linewidth}
        \centering
        \includegraphics[width=\linewidth]{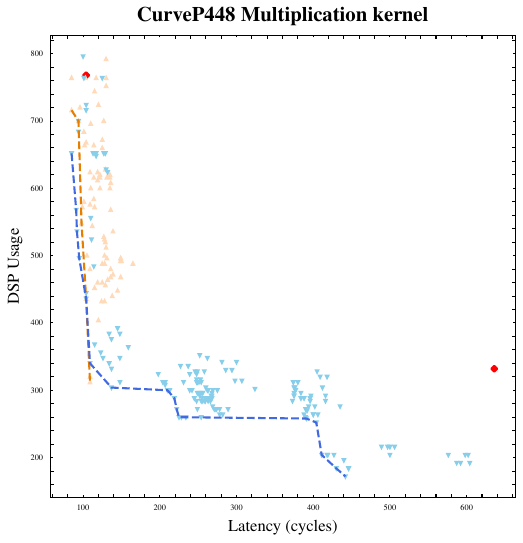}
        \caption{Design Points for CurveP448 Multiplication Kernel}
        \label{fig:paretoC}
    \end{subfigure}
    \hspace{0.05\linewidth} 
    \begin{subfigure}[t]{0.41\linewidth}
        \centering
        \includegraphics[width=\linewidth]{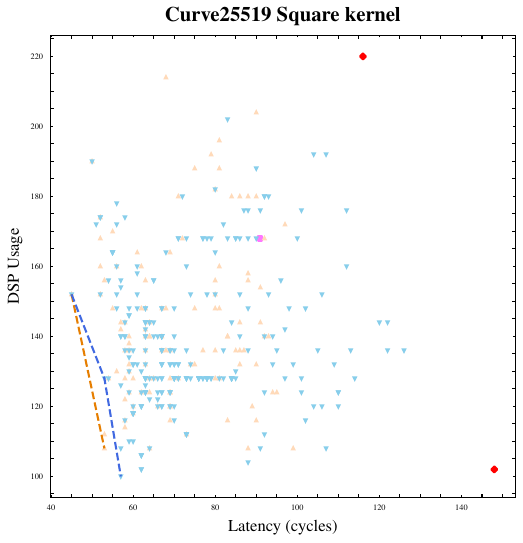}
        \caption{Design Points for Curve25519 Square Kernel}
        \label{fig:paretoD}
    \end{subfigure}
    
    \caption{Pareto curves obtained from DSE on single cryptographic kernels.}
    \label{fig:DSE}
    \vspace{-15pt}
\end{figure*}

\subsubsection{QoR Estimator} 
A key factor driving Cryptonite's performance is its Design Space Exploration (DSE) combined with Iterative QoR-Guided Optimization. The Quality of Results (QoR) estimator is used to predict and evaluate the effectiveness of a hardware design in terms of key performance metrics before synthesis. The QoR estimator includes an analytical model that predicts the latency (in cycles) and DSP resource usage based on loop structure, dependencies, and applied optimizations. Accurately predicting resource usage is challenging as it depends on factors such as routing and overall design complexity. The explorer initially generates all feasible designs. The QoR estimator compares designs, eliminates dominated solutions (i.e., non-Pareto solutions), and retains the optimal Pareto designs. The Pareto frontier designs are given as feedback (QoR feedback) to apply additional optimizations to refine the Pareto front to enable the exploration of new points near the Pareto front.



\section{Experimental Results}
\label{sec:Results}


Cryptonite is evaluated by conducting experiments leveraging the straight-line primitives from the Fiat Project. The primitives included in the results are Elliptic Curves \texttt{25519}, \texttt{P521}, \texttt{P448}, \texttt{Secp256k1}, and \texttt{P384}. \texttt{Curve 25519} enables efficient Diffie-Hellman key exchange. \texttt{Curve P521}, offers higher security with increased resource consumption. \texttt{Curve P448} provides a balance between security and performance compared to \texttt{Curve P521}. \texttt{Curve Secp256k1} is used for secure key-pair generation and digital signatures. Finally, \texttt{Curve P384} offers superior security over \texttt{Curve Secp256k1}, using more resources.

Cryptonite's DSE results are compared to manual optimization and Scale-HLS, with Cryptonite achieving more scalable designs compared to manual rewriting. The results obtained from the manual rewriting are synthesized using the Vitis HLS tool 2023.1 version targeting the \texttt{xczu9eg-ffvb1156-2-e} FPGA board. While generating hardware-friendly loop structures is challenging, handwritten RTL benchmarks often outperform the RTL generated by the Vitis HLS compiler. Creating such benchmarks is beyond the scope of this work due to the extensive development time required. When these primitives are synthesized on hardware, the lack of structural variation extends to the RTL representation limiting the effectiveness of comparison between baseline code and the HLS generated design. Instead of focusing on an RTL-to-RTL comparison, we evaluate performance metrics such as latency and area to explore optimizations that balance efficiency and correctness rather than adhere to an undefined ground truth RTL. 

This work primarily focuses on accelerating low-level cryptographic primitives such as modular multiplication that are widely used in many cryptography applications such as elliptic curves and lattice-based cryptography. This work leverages elliptic curve cryptography to illustrate the approach, but none of the presented methods are specifically tailored to elliptic curve cryptography. The approach is designed to work for any straight-line C-code application, provided there are opportunities for loop parallelization to enable re-rolling.

We observed that as a program is optimized to utilize fewer DSPs, it becomes more generalized, leading to an increase in the usage of LUTs. Since DSP blocks are specialized for handling complex operations, it is important to conserve them to enable the concurrent execution of different applications on the FPGA.

Cryptonite is tested for 12 distinct primitive types in the Fiat Cryptography project~\cite{fiat19}. For brevity, we provide a graphical comparison of the scalability achieved by manual rewriting, Scale-HLS, and Cryptonite for four cryptographic primitives, considering both single and multi-kernel implementations. The results are presented in \hyperref[tab:table]{Table~\ref*{tab:table}} and \hyperref[fig:DSE]{Fig.~\ref*{fig:DSE}}  \& \hyperref[fig:MultiDSE]{Fig.~\ref*{fig:MultiDSE}} respectively. This section is further divided into two parts. \hyperref[sec:results-single]{Sec.~\ref*{sec:results-single}}  demonstrates the \textbf{impact of a single kernel implementation}, while  \hyperref[sec:results-multiple]{Sec.~\ref*{sec:results-multiple}} explores about the \textbf{impact of multiple kernel implementations}. ScaleHLS was unable to produce synthesizable designs for all input cryptographic primitives. For these cases, the ScaleHLS results are omitted in \hyperref[tab:table]{Table~\ref*{tab:table}} and \hyperref[fig:paretoD]{Fig.~\ref*{fig:paretoD}}. The synthesis results for the manually optimized and Cryptonite designs are still provided for comparison.

\subsection{DSE on Single Cryptographic Kernel}
\label{sec:results-single}
\hyperref[tab:table]{Table~\ref*{tab:table}} summarizes the synthesis results from implementing a single kernel for the cryptographic primitives discussed earlier. Cryptonite demonstrated up to a 2.75$\times$ speedup and a 2.03$\times$ reduction in resource usage. Notably, the tool identified design points with lower resource usage than manual optimizations and Scale-HLS. Cryptonite also achieves the lowest latency design for primitives other than \texttt{Curve secp256k1} and \texttt{Curve P384}. These results stem from the dependency of the DSE engine on Loop Synthesis. When the e-graph stage struggles to generate effective loop structures, there are limited opportunities to apply latency-oriented optimizations automatically. The transformations in primitives \texttt{Curve secp256k1} and \texttt{Curve P384} outline repeated calls to a common arithmetic function into a shared master function to improve modularity and enable code reuse. This approach naturally reduced resource utilization by allowing time-multiplexed reuse of DSPs. However, these operations could not be pipelined due to loop-carried dependencies introduced by carry propagation chains. Although the transformation reduced resource usage significantly, the overall latency increased compared to the original straight-line code.

\hyperref[fig:paretoA]{Figs.~\ref*{fig:paretoA}}--\hyperref[fig:paretoD]{~\ref*{fig:paretoD}} present the Pareto curves corresponding to synthesis outcomes for four distinct kernels. The plotted points represent design candidates explored using Cryptonite, ScaleHLS, and manual tuning of design parameters. The automated DSE can often generate the best set of Pareto Optimal designs. In three of the four kernels, designs discovered by Cryptonite dominate both the naive straight-line C and fully-rolled implementations in latency and resource usage. Additionally, designs discovered by Scale-HLS only extend the Pareto-curves produced by Cryptonite, re-emphasizing the optimality of the designs generated automatically.

These results highlight the effectiveness of the loop-synthesis stage in enabling practical design space exploration. All three methods (manual, automated, and Scale-HLS) produce a range of superior hardware designs to direct straight-line C implementations. Furthermore, Cryptonite is frequently able to automatically generate a set of Pareto-optimal designs with a range of resource-latency trade-offs, which is competitive with both manual optimization and ScaleHLS. 

\subsection{DSE on Multiple Cryptographic Kernels}
\label{sec:results-multiple}

\begin{figure}[h!]
    \centering
    \begin{subfigure}[t]{1.0\linewidth} 
        \centering
        \includegraphics[width=\linewidth, height=0.65\linewidth]{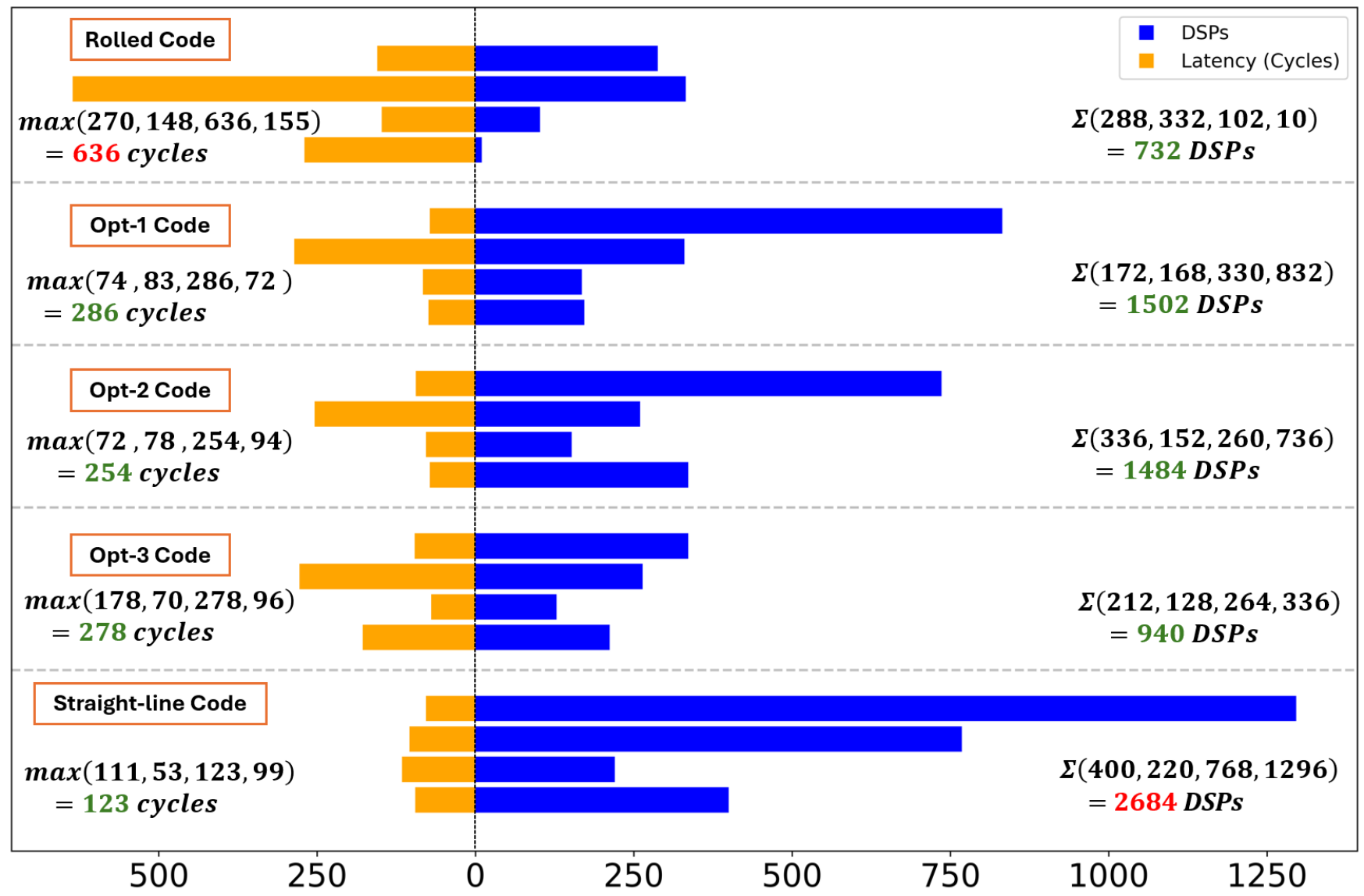}
        \caption{Cryptonite's Multi-kernel DSE}
        \label{fig:MultiDSE1}
    \end{subfigure}
    
    \vspace{0.2cm} 

    \begin{subfigure}[h!]{1.0\linewidth} 
        \centering
        \includegraphics[width=1\linewidth, height=0.5\linewidth]{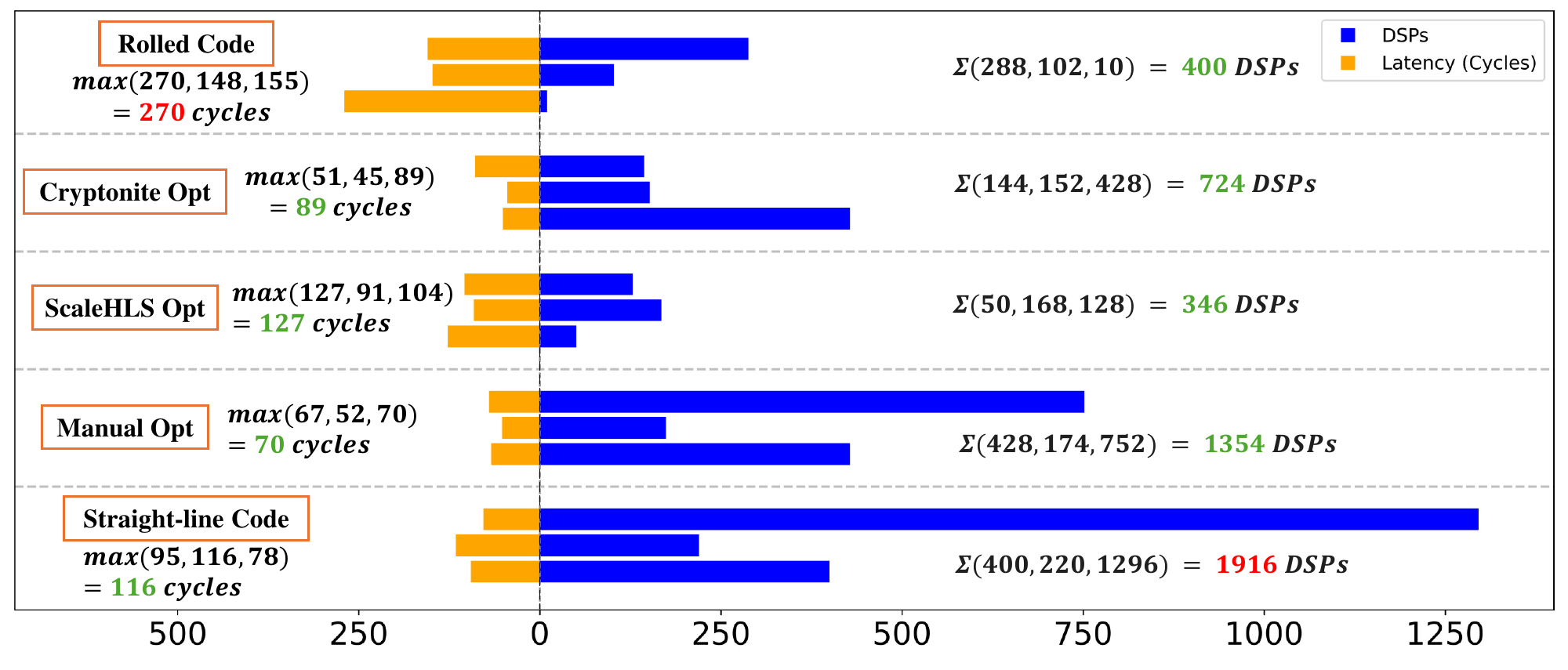}
        \caption{Comparison Multi-DSE results for Cryptonite, Scale-HLS and manual Rewriting}
        \label{fig:MultiDSE2}
    \end{subfigure}
    \caption{DSE on multiple cryptographic kernels.}
    \label{fig:MultiDSE}
\end{figure}

Cryptonite concurrently merges Pareto optimal implementations of multiple kernels, for joint deployment on a resource-constrained FPGA board. \hyperref[fig:MultiDSE]{Fig.~\ref*{fig:MultiDSE}} compares the outputs of different execution strategies, including straight-line C-codes, Pareto-optimal designs, and loop-synthesized codes, when multiple kernels are executed together. Total consumption of resources is given by $\sum_{i=1}^{N} D_i$, where $D_i$ is the resource count for a kernel in the $i^{th}$ permutation and the effective latency for that permutation is $max(L_0, L_1, ...., L_i)$, where $L_i$ is the Latency of a $i^{th}$ Kernel. In \hyperref[fig:MultiDSE1]{Fig.~\ref*{fig:MultiDSE1}}, the design points opt-1, opt-2, and opt-3 utilize 59.6\%, 58.88\%, and 37.3\% of the total DSPs (2520) available on the target FPGA, respectively. \hyperref[fig:MultiDSE2]{Fig.~\ref*{fig:MultiDSE2}} compares the multi-kernel DSE results obtained for Manual, Scale-HLS and Cryptonite optimizations for the kernels \texttt{Curve25519 Mul}, \texttt{Curve25519 Square} and \texttt{CurveP521 Mul} respectively. Compared to naively synthesized straight-line code, Cryptonite reduces DSP usage by approximately 62.21\%, while also reducing execution time by 27 cycles. Scale-HLS, on the other hand, achieves a 81.94\% reduction in DSP usage, but results in a 11-cycle increase in total execution time. Meanwhile, the manual optimization reduces DSP usage by 29.33\% and outperforms the straight-line code by 46 cycles in terms of execution time. Designs synthesized using naive straight-line C implementations of multiple kernels are unable to achieve the same resource utilization, highlighting Cryptonite's ability to generate scalable designs automatically. 

\section{Related Work}
\label{sec:RelatedWork}

The Fiat Cryptography project~\cite{fiat19} automatically generates code for cryptographic primitives based on proofs in the interactive theorem prover \textbf{Coq}. It outputs primitives for several curves that are provably correct in assembly and several high-level languages, including C. We target the code generated in C using our approach. The fact that this code is auto-generated also creates many opportunities for pattern detection. These patterns can be used to automatically try different ways of re-structuring the code.

Previous works have proposed manually optimized architectures for cryptographic primitives, some of them particularly focused on improving efficiency to achieve the best performance~\cite{ernst04, chelton08}. Others have focused on manually designing accelerators for particular constrained settings, like Internet of Things (IoT) devices~\cite{laranino20, sandoval21}.

SEER \cite{seer23} serves as a super-optimization explorer for HLS, utilizing e-graphs to implement rewrite rules. However, this work focuses on applying MLIR-base optimization rather than loop generation for HLS DSE.  Furthermore, the MLIR-based optimizations are not well-suited to cryptography and concentrate solely on latency minimization. Nandi \emph{et al.}~\cite{nandicad} also uses e-graphs for loop re-rolling in the context of shrinking de-compiled computer-aided design (CAD) models. Since this work targets de-compiled code, the synthesized loops likely existed in the original program and were then unrolled. Since cryptographic primitives do not initially include loops, the source code must be transformed to eliminate local variables and expose patterns that can be restructured into loopable code.


ScaleHLS~\cite{ye2021scalehls} is a tool leveraging MLIR to transform and optimize C/C++ code for High-Level Synthesis (HLS) at various abstraction levels. While it excels at enhancing the performance of kernels used in machine learning workloads, it is not well-suited for handling the looped codes generated from the straight-line codes produced by the Fiat Project. AutoScaleDSE~\cite{10.1145/3572959}, an extension of ScaleHLS tailored for ML workloads, automates and optimizes design space exploration (DSE) to achieve efficient scaling of compute-intensive systems by balancing performance, resource utilization, and constraints. Similarly, the MLIR base optimizer, relying on Affine and SCF dialects, struggles to find meaningful cryptographic optimization before and after loop \& array synthesis. 

POLSCA~\cite{polsca22} is another non-MLIR optimization framework, focusing on source code optimization with HLS-friendly polyhedral transformations. However, it also relies on structured loops and affine operations, making it unsuitable for cryptographic primitives. Additionally, unlike prior works~\cite{seer23, ye2021scalehls, polsca22}, which primarily target performance speedup, our work emphasizes achieving scalable designs to meet broader application requirements.



While side-channel resistance is not the primary focus of this work, we recognize its critical importance when implementing cryptographic protocols. Though our source code transformation does introduce branches, these statements never branch on secret values, but rather on statically determined constants that determine loop length. However, these branches will not introduce side-channel vulnerabilities at the software level. The aforementioned hardware implementation explorer ensures that critical properties of the original code, particularly those related to constant-time arithmetic, manual carry propagation, ambiguity in bit widths, and lack of variable loop bounds, are preserved. It avoids introducing control-flow variability, operator widening, or replacing masking logic with more complex arithmetic. By retaining these properties, the transformations applied by the explorer remain potentially side-channel resistant, enabling safe application of HLS optimizations.

Some side-channel vulnerabilities are unique to FPGA-based cryptographic implementations. Prior work, such as~\cite{10.1145/3492345}, highlights how automatic hardware generation via HLS can inadvertently introduce timing and resource usage patterns that leak secret information. These findings are outside of the scope of our work; our approach assumes that the HLS tool will not introduce new side-channel vulnerabilities. MaskedHLS~\cite{sarma2024maskedhlsdomainspecifichighlevelsynthesis} presents a domain-specific HLS framework for synthesizing masked side-channel-resistant designs. This illustrates that proper tool-chain support can mitigate these risks.

Researchers have done significant work on hardware acceleration for cryptography protocols. Devadas et. al \cite{devadas22} explores the benefits and challenges in this domain. This work concludes that hardware acceleration is very effective for cryptography. Although developing custom hardware for one protocol usually produces the fastest results, this may not be the best option due to the ever-evolving nature of the encryption standard. Recent works investigate accelerating post-quantum cryptography (PQC) workloads. One work~\cite{agrawal19} proposes an open-source FPGA implementation of PQC primitives. Unlike our work, these are manually implemented and optimized for each primitive. Another suggests designing a ring processing unit (RPU)~\cite{soni23} with a specialized ISA and microarchitecture tailored to Ring-LWE PQC protocols. Our work improves automatic design generation from a high-level specification.


\section{Conclusion}
\label{sec:Conclusion}
This paper introduces Cryptonite, a tool designed to scale hardware implementations of cryptographic primitives in resource-constrained environments. By leveraging source code transformations in combination with HLS feedback, Cryptonite explores the design space of cryptographic primitives, generating diverse hardware design points. This flexibility enables the selection of optimal designs based on specific user constraints. The results from its multi-kernel implementation pave the way for more efficient hardware acceleration for cryptographic primitives.
\section*{Acknowledgment}
The authors thank the anonymous reviewers for their feedback on earlier drafts of this paper. The work described in this paper was supported, in part,  by the United States National Science Foundation (NSF) under grants No. 2114627, No. 2237440 and No. 2317251. Any opinions, findings, conclusions, or recommendations expressed in this publication are those of the authors and do not necessarily reflect the views of the above sponsoring entities.

\begingroup
\tiny
\bibliographystyle{ieeetr}

\endgroup

\end{document}